\documentclass[11pt,a4paper]{article}
\usepackage{jheppub,amsmath,  amssymb,slashed,url,bm}
\usepackage{amsfonts}
\usepackage{graphicx}
\usepackage{epstopdf}

\title{\center{  \center{$n$-point correlations of dark matter tracers}  \\  ~\\  \small{\center{Renormalization with univariate biasing} \\ \center{and} \\ \center{its $O(f_{NL})$ terms with bivariate biasing}} } } 

 \author{Anirbit}
\affiliation{Physics Illinois,\\University of Illinois at Urbana-Champaign (UIUC)\\}
\emailAdd{amukher4@illinois.edu}
\abstract
{ 
In this note we extend the results of the model of galaxy biasing recently proposed in \cite{FDV}. In that paper the authors had outlined a very precise mathematical framework to deal with the theory of galaxy biasing. In this work we extend that combinatorial technology to renormalize tracer $n-$point functions (\ref{renbiasuni}) in their model of univariate biasing. We further prove that $4$ and higher point cumulants of the Bardeen potential don't have an $O(f_{NL})$ term, (Appendix \ref{fnl}). Then we use this observation to extract all the $O(f_{NL})$ terms in the $n-$point correlation of tracers in their model of bivariate biasing, (\ref{final}). 
}
\begin{document} \maketitle

\section{Introduction}\label{intro} 

A postulated inflationary phase of the universe at the beginning still remains as the most viable bet to explain many of the mysterious aspects of the universe like the near uniformity (hence correlation) of the CMB temperature on scales even beyond the causal horizon. (\cite{WMAP9}). There are many well-developed and sophisticated theories to produce and study the inflationary universe (\cite{BF}, \cite{Sandip}, \cite{Randall}, \cite{Creminelli}, \cite{Khoury}, \cite{B}, \cite{Jessie})\\
~\\
At a crude level the different models of inflation differ in $(1)$ their values of the spectral index (or the spectral tilt from the slow-roll single-field model of $n_s=1$), $(2)$ the amount of gravitational waves from the early universe, $(3)$ isocurvature perturbations (from the multi-field models), $(4)$ non-Gaussianity of the primordial inflaton field. In this work we are mainly concerned with the possible non-Gaussianity of the inflaton field.\\
~\\ 
In this work we restrict ourselves to the linear regime, that is times much after the equilibrium epoch and length-scales well within the causal horizon. In this regime the $k^{th}$ Fourier mode of the matter perturbation, $\delta_k$ is proportional to the $k^{th}$ Fourier component of the primordial gravitational potential, $\Phi_{kp}$. (using the definitions in equations $7.5$ and $7.8$ of (\cite{Dodelson}) one has the relationship $\delta_{kp} = \frac{3}{5\Omega_m H_0^2}(k^2T(k)D(a))\Phi_{kp}$). {\emph Though we emphasize that the exact value or form of the proportionality constant is not going to be relevant anywhere in this work.} Also instead of working in the momentum space it will be convenient for us to be working in the real space.\\
~\\
In the recent past, (\cite{NGLS}) it has been realized that at least this basic level of primordial non-Gaussianity will be far better probed than the CMBR by the galaxy data that is likely to be accumulated by the upcoming large scale surveys like DES, BOSS, LSST.\\
~\\
But what these surveys detect are some ``tracers" (like QSOs or Luminous Red Galaxies (LRGs)) and not the dark matter directly. But its the dark matter distribution which is theoretically most accessible since it being the predominant ``thing" affects the gravitational potential the most. Hence one needs a theoretical model which relates the detectable matter to the dark matter halos. We believe that all matter in the universe resides inside huge dark matter halos. The galaxies are supposed to form inside these halos and that produces a ``bias" between the kind of matter distribution (like type of galaxy) detected and the dark matter density. There is little or no theoretical understanding of this phenomenon of ``biasing" but till date the simplest model of ``local biasing" (as defined in the next section) has proven to be largely useful. Thus we build on a long history of papers on biasing theory \cite{Sarah}, \cite{D}, \cite{Baldauf}, \cite{others}.\\
~\\
In the biasing model of \cite{FDV} it becomes possible to ask as to what is the imprint of primordial non-Gaussianity on the large-scale structure of the universe. Within the biasing model of \cite{FDV} we prove a set of mathematical results namely, a proof of renormalizability of tracer n-point functions with univariate biasing (\ref{renbiasuni}) and a theorem about cumulants (\ref{fnl}) which we will use to determine the $O(f_{NL})$ contribution to the tracer n-point function in the model of bivariate biasing, (\ref{final}). 

In appendix \ref{universal} we make some observations about possible ``nzl" decompositions in the model of bivariate biasing and in appendix \ref{y} we make some comments about the idea of the ``y" parameter of \cite{FDV}.

\newpage

\section {Bias parameters and their need to be renormalized}

We work in the model whereby the dark matter halos are assumed to be housing the galaxies and hence mapping out the galaxies gives information about the former. In such a case we call the dark matter halos or the galaxies as ``tracers" for the matter distribution. We consider the matter distribution to mean all of matter including unknown (dark matter) and the known (the standard model).  Conventionally this tracing relationship is stated as a power-series of the form, 

\begin{align}
\langle n_{h}(\vec{x}) \rangle = \sum _ {i=0}^{\infty} \frac{c_i}{i!} \langle \delta_{L}^i(\vec{x}) \rangle 
\end{align} 
 
Here $n_h(\vec{x})$ is the number density of dark matter tracers at the position $\vec{x}$ and $\delta_L (\vec{x})$ is the smoothed fractional overdensity at spatial position $\vec{x}$ of the matter density field. For any density field $\rho$ the fractional overdensity is defined by $\delta = \frac{\delta \rho}{\rho}$ and here the $\delta$'s are necessarily smoothed on a scale $R_L$. The subscript $L$ on any variable denotes being in ``Lagrangian space" and this will henceforth be dropped for notational convenience. The coefficients $c_i$ in the above power-series expansion are called ``bias parameters" (or in anticipation of what is to come we more specifically call it the ``bare bias parameters"). \\
~\\

One eventually wants to consider the n-point functions of these tracers ($\xi_h$) for any given set of $n-$points say $\{\vec{x}_i\}_{i=1}^{i=n}$ as,

\begin{align}
\xi_h( \vert \vec{x}_i - \vec{x}_j \vert_{(0<i<j=2,..,n)} ) = -1 + \frac{\langle \prod_{i=1}^{i=n} n_h(\vec{x}_i) \rangle  } {\prod_{i=1}^n \langle n_h(\vec{x}_i) \rangle} 
\end{align}
 
In the above we ensemble average over all the variables on which the functions depend on and hence in particular, $\langle n_h(\vec{x}_i)\rangle = \langle n_h(\vec{x}_j)\rangle$ for any two $i$ and $j$.  Henceforth we denote $\delta_L(\vec{x}_i)$ as just $\delta_i$ 

One now sees that the substitution of the first equation into the second will not lead to any meaningful notion of $n-$point functions because of the following two issues, 

\begin{itemize}
\item Firstly the $\xi_h$ thus defined would depend on the unphysical parameter - the smoothing scale $R_L$ of $\delta_L$ and that can't be allowed to happen.  

\item Secondly the correlation function thus defined would involve terms of the kind $\langle \delta_i^n \rangle$ which are firstly potentially divergent and also they don't drop to zero when the points are taken far away from each other. This implies that quite catastrophically even for infinitely separated $n$-points the power-series is not guaranteed to converge because none of the terms are guaranteed to be small. 
\end{itemize}

We will see in this note that an objective accomplished will be that we shall be able to rearrange the above expression such that it looks like a sum of products such that each of the products cleanly splits into a $R_L$-independent part and a factor which decays off when any two points are taken to infinite separation. 

\section {Rearranging the expression for the $n$-point function} 

Let us believe that there exists a function $F_{h,L}(\delta,\vec{x})$ such that $n_h(\vec{x}) =  F_{h,L}(\delta,\vec{x})$. Henceforth we shall drop the subscript $L$ on this $F$ function as well.  Now doing a Taylor series expansion of the of the function $F_h$ on its variable $\delta$ we have for each point $\vec{x}_i$,

\begin{align}
n_h(\vec{x}_i) = \sum _ {n_i =0} ^ {\infty} \frac{F^{(n_i)}_h(0,\vec{x}_i)}{n_i !} \delta_{i}^{n_i}
\end{align}

One can now see the $c_i$s as being the same as the $\langle F^{(n_i)}(0,\vec{x}_i) \rangle$. In what follows we shall see that for understanding some of the assumptions needed for the analysis, its more illuminating to think in terms of $\langle F^{(n_i)}(0,\vec{x}_i) \rangle$ rather than the $c_i$s.  Now substituting this into the previous expression for the n-point function one has, 

\begin{align}
\langle \xi_h(\vert \vec{x}_i - \vec{x}_j\vert_{(0<i<j=2,...,n)})\rangle = -1 + \frac{ \sum _{n_{i=1...n} =0} ^{n_{i=1...n} =\infty }  \langle  \prod_{i=1}^{n} \left [  \frac{1}{n_i !} F_h^{(n_i)}(0;\vec{x}_i) \delta^{n_i}_i   \right ] \rangle }   { \prod _{i=1}^{n} \left [  \sum_{n_i =0} ^{\infty} \frac{1}{n_i !} \langle F_h^{(n_i)}   (0;\vec{x}_i) \delta ^{n_i}_i \rangle \right ] } 
\end{align} 

Now we assume that  $F_h^{m}(0,\vec{x})$ and $\delta ^m (\vec{x})$ are uncorrelated variables at every point $\vec{x}$. Then it follows that, 

\begin{align}\label{Fas1}
\langle  \prod_{i=1}^{n} \left [  F_h^{(n_i)}(0;\vec{x}_i) \delta^{n_i}_i \right ] \rangle = \langle \prod_{i=1}^{n} F_h^{(n_i)}(0;\vec{x}_i) \rangle \langle \prod_{i=1}^{n} \delta^{n_i}_i \rangle 
\end{align}  

So this allows the $n$-point function to be rewritten as,

\begin{align}\label{npoint} 
\langle \xi_h(\vert \vec{x}_i - \vec{x}_j\vert_{(0<i<j=2,...,n)})\rangle = -1 + \frac{ \sum _{n_{i=1...n} =0} ^{n_{i=1...n} =\infty }  \langle  \prod_{i=1}^{n} \left [  \frac{1}{n_i !} F_h^{(n_i)}(0;\vec{x}_i) \right ] \rangle \langle \prod_{i=1}^n \delta^{n_i}_i   \rangle }   { \prod _{i=1}^{n} \left [  \sum_{n_i =0} ^{\infty} \frac{1}{n_i !} \langle F_h^{(n_i)}   (0;\vec{x}_i) \rangle \langle \delta ^{n_i}_i \rangle \right ] } 
\end{align}

\subsection {Defining non-zero-lag correlators} 

Firstly one wants to rewrite the correlation function  $\langle \prod_{i=1}^n \delta^{n_i}_i   \rangle$ into a sum of product of cumulants. In the language of QFT, this is the same as decomposing a correlation functions into a sum of product of connected correlation functions. The $n-$ point correlation functions are the $n-$fold derivative of the partition function with respect to the source currents corresponding to the $n-$ fields in the correlation function and the connected correlation functions are the same differentiation done on the logarithm of the partition function. We present here a way of writing this relationship. \\
~\\
Imagine a tableau which looks like the Young's tableau (but not that!) such that each of the rows are labeled by $\delta_i$ and each row has $n_i$ boxes for $i=1$ to $i=n$. Let $\Pi_{n_{i=1...n}}$ be the set of all partitions of this tableau. Then let $\rho$ label the elements of $\Pi_{n_{i=1...n}}$. For any $\rho$ let $B$ label the subsets or the parts in the partition. And let $n_i(B)$ be the number of elements of $B$ which come from the row labeled by $\delta_i$.    With these set of definitions the following relation follows for expressing any $n-$point function into a sum of products of connected correlation functions (labeled with a subscript $c$),

\begin{align}
\langle \prod_{i=1}^n \delta^{n_i}_i   \rangle = \sum _{\rho \in \Pi_{n_{i=1...n}}} \prod _ {B \in \rho} \langle \prod_{i=1}^n \delta_i^{n_i(B)} \rangle_c  
\end{align} 

Now define the ``non-zero-lag" (``nzl") correlation ($\langle \prod_{i=1}^n \delta^{n_i}_i   \rangle_{nzl}$) as that part of the RHS of the above where $\rho$ are such that it has no $B$ which is contained wholly in any single row (corresponding to some $\delta_i$). So ``nzl" correlation removes the contributions from all self-moments of the field at any one point and hence its guaranteed to go to $0$ as any two of the points is taken to infinite separation.  With this definition one can show that (Appendix \ref{nzl}) the following is true,  

\begin{align}\label{nzl} 
\langle \prod_{i=1}^n \delta^{n_i}_i   \rangle = \sum _ {k_{i=1...n}=0} ^{k_{i=1...n}=n_i}  \prod_{j=1}^{n} \left [ ^{n_j}C_{k_j} \langle \delta _j ^{k_j} \rangle \right ]   \langle  \prod_{l=1}^{n} \delta_l^{n_l - k_l}  \rangle_{nzl} 
\end{align} 

Henceforth we shall often use this notation of $^{n}C_k = \frac{n!}{k!(n-k)!}$.

\subsection {Defining peak-bias-split (PBS) bias parameters} 

The essential hypothesis of the peak-bias-split formalism is that $\langle n_h \rangle$ which is initially defined in terms of matter overdensity $\delta$ can be rewritten in terms of the the absolute density ($\rho$) such that the following is true for any overdensity $\delta = D$ defined about a background density $\bar{\rho}$ 

\begin{align}
\langle n_h \rangle (\rho = \bar{\rho}(1+D)) = <n_h> (\delta = D) 
\end{align} 

So it follows that if one starts with an overdensity say $\delta$ (and hence a density of $\rho = \bar{\rho}(1+\delta)$) and then uniformly shifts the density by $D\bar{\rho}$ then the density changes to $\bar{\rho}(1 + \delta + D)$. So an uniform density shift of $D\bar{\rho}$ is equivalent to changing the overdensity from $\delta$ to $\delta + D$.  Hence about a fixed background density $\bar{\rho}$ the following equality will hold at any point $\vec{x}_i$,

\begin{align}
\langle n_h (\vec{x}_i) \rangle (D)  = \sum _{n_i =0} ^{\infty} \frac { \langle F^{(n_i)} _ h (0;\vec{x}_i) \rangle  } {n_i ! } \langle (\delta_i + D )^{n_i} \rangle     =  \sum _{n_i =0} ^{\infty} \frac {\langle F^{(n_i)} _ h (0;\vec{x}_i) \rangle  } {n_i ! } \sum_{r_{i} =0}^{n_i} \text{ } ^{n_i}C_{r_i} \langle \delta _i ^{r_i} D^{n_i - r_i} \rangle 
\end{align}

From the above it follows that,  

\begin{align}\label{bn}
b_N := \frac{1}{\langle n_h(\vec{x}_i) \rangle (D=0)} \frac{\partial ^N \langle n_h (\vec{x}_i )\rangle (D)} {\partial D^N} \vert _{D=0} = N! \frac { \sum _{n_i = N} ^{\infty} \frac{\langle F_h ^{(n_i)} (0,\vec{x}_i) \rangle  } { n_i !} ^{n_i}C_{n_i - N} \langle \delta _i ^{n_i - N} \rangle }      { \sum_{n_i = 0} ^{\infty}  \frac{\langle F_h ^{(n_i)} (0,\vec{x}_i) \rangle  } { n_i !}  \langle \delta _i ^{n_i} \rangle  }   
\end{align} 

Given the PBS framework, these newly defined variables $b_N$ can equivalently be defined as $b_N = \frac{\bar{\rho}^N}{\langle n_h \rangle} \frac{\partial ^N \langle n_h \rangle }{\partial \bar{\rho}^N}$. This relationship implies that $b_N$ can be defined wholly in terms of the average background density ($\bar{\rho}$) (as opposed to the overdensity) and the number density of tracers. (which is also a function of the average background density) None of these need a smoothing scale to be defined and hence its legitimate to call $b_N$ the ``renormalized" bias parameters.

\subsection {The proof of renormalizability of $n-$point functions for the case of univariate biasing} 

Now substituting the definition of the ``nzl" correlations as in \ref{nzlfinal} in the expression for the $n-$point function as in \ref{npoint}  we have, 

\begin{align}
\notag &\langle \xi_h(\vert \vec{x}_i - \vec{x}_j\vert_{(0<i<j=2,...,n)})\rangle\\
 &= -1 + \frac{ \sum _{n_{i=1...n} =0} ^{n_{i=1...n} =\infty }  \langle  \prod_{i=1}^{n}   \frac{1}{n_i !} F_h^{(n_i)}(0;\vec{x}_i) \rangle \sum _ {k_{i=1...n}=0} ^{k_{i=1...n}=n_i}  \prod_{i=1}^{n} \left [ ^{n_i}C_{k_i} \langle \delta _i ^{k_i} \rangle \right ]   \langle  \prod_{i=1}^{n} \delta_i^{n_i - k_i}  \rangle_{nzl}   }   { \prod _{i=1}^{n} \left [  \sum_{n_i =0} ^{\infty} \frac{1}{n_i !} \langle F_h^{(n_i)}   (0;\vec{x}_i) \rangle \langle \delta ^{n_i}_i \rangle \right ] } 
\end{align} 

Now define a new set of variables $N_i = n_i - k_i >0$ and we have, 

\begin{align}\label{npoint1}
\notag &\langle \xi_h(\vert \vec{x}_i - \vec{x}_j\vert_{(0<i<j=2,..,n)})\rangle \\
&= -1 + \frac{ \sum_{N_i = 0} ^ {\infty} \sum _{n_i = N_i, i = 1...n} ^{n_{i=1...n} =\infty }  \langle  \prod_{i=1}^{n} \frac{1}{n_i !} F_h^{(n_i)}(0;\vec{x}_i) \rangle \prod_{i=1}^{n} \left [ ^{n_i}C_{n_i - N_i} \langle \delta _i ^{n_i - N_i} \rangle \right ]   \langle  \prod_{i=1}^{n} \delta_i^{N_i}  \rangle_{nzl}   }   { \prod _{i=1}^{n} \left [  \sum_{n_i =0} ^{\infty} \frac{1}{n_i !} \langle F_h^{(n_i)}   (0;\vec{x}_i) \rangle \langle \delta ^{n_i}_i \rangle \right ] } 
\end{align} 

Now we make the assumption that arbitrary order overdensity derivatives of the function $F_h$ at different points are uncorrelated. Which means, 

\begin{align}\label{Fas2}
\langle \prod_{i=1}^{n} F_h^{(n_i)}(0;\vec{x}_i) \rangle = \prod _{i=1}^{n_i} \langle F_h ^{(n_i)}(0;\vec{x}_i)\rangle 
\end{align}

With this last assumption, one can replace the definition of $b_N$ (\ref{bn}) into the expression for $\xi_h$ (\ref{npoint1}) to get the final expression, 

\begin{align}\label{renbiasuni}
\langle \xi_h(\vert \vec{x}_i - \vec{x}_j\vert_{(0<i<j=2,..,n)})\rangle = \sum_ {N_{i=1...n}=0} ^ {\infty} \left [  \prod_{i=1}^{n} \frac{b_{N_i} }{ N_i !} \right ] \langle  \prod_{i=1}^n \delta _i ^{N_i} \rangle _{nzl}
\end{align} 

Thus one as rewritten the $n-$point function as a sum of product of terms whereby in each product one factor (the $b_N$s) are independent of the coarse-graining scale and the the other factor (the ``nzl" correlations) is such that it goes to zero when any two points are infinitely separated. Thus this rewriting makes more physical sense and resolves both the issues of coarse-graining dependence and uncontrolled divergence of the initial expression.


\section {$O(f_{NL})$ terms in bivariate biasing} \label{fNLn}

In the appendix (\ref{y}) we have reviewed the idea of \cite{FDV} of introducing a variable $y$ to characterize bivariate biasing. Since any occurrence of the variable $y$ in the correlation function is effectively an occurrence of $2$ $\delta$'s this helps in keeping track of the powers of $f_{NL}$ in the power-series expansion of the tracer $n-$point function. In this section we shall formalize this idea and derive an exact result in this model for all the $O(f_{NL})$ terms in the tracer $n-$point function.\\
~\\
Doing the similar analysis as in the previous sections for the case of bivariate biasing we have for some numbers $c_{n,m}= F^{(n,m)}(0,0,\vec{x}_i)$ ,

\begin{align} \label{bin}
 \langle \xi (\vert  \vec{x}_i - \vec{x}_j \vert_{(0<i<j=2,...,n)}) \rangle = -1 + \frac {\sum_{\{ n_i,m_i =0\}_{i=1}^n} ^\infty     \prod _{i=1}^n \left [  \frac{c_{n_i,m_i} }{n_i! m_i !} \right ] \langle  \prod _{i=1}^n \delta _i ^{n_i} y_i^{m_i} \rangle }
{\prod _{i=1}^n \left [ \sum_{\{ n,m =0\}} ^\infty \left [  \frac{c_{n,m} }{n! m!} \right ]   \langle \delta ^{n} y^{m} \rangle \right ] }  
\end{align}

(..by the assumption of spatial homogeneity the subscript $i$ has been removed from the fields appearing in the correlations in the denominator..) Now in the above expression we would want to collect together all terms that contribute at $O(f_{NL})$. But the correlations appearing in the numerator and the denominator of the fraction on the RHS of the above equation are not cumulants/connected correlations. Hence for any given correlation function there are very complicated ways in which the constituent cumulants will combine to produce $O(f_{NL})$ terms in its expansion. Any correlation function occurring in the fraction will have a cumulant expansion and we focus on the terms occurring in the cumulant expansion to try to track down the required shapes of product of cumulants.\\
~\\
In the linear regime where we are working we have $ \delta \sim \nabla \varphi$ where $\varphi$ is the Bardeen potential and we are parameterizing non-Gaussianity as $\varphi = \phi + f_{NL}(\phi^2 - \langle \phi^2 \rangle)$. Now if we are to calculate some $n-$point cumulant of the overdensity $\delta$ then we have $\langle \prod _{i=1}^n \delta _i \rangle \sim \langle \prod_{i=1}^n \nabla _i \varphi \rangle $. The $\nabla$ has a subscript $i$ to denote that $\varphi_i = \varphi(\vec{x}_i)$ is being differentiated w.r.t the coordinates there (i.e $\vec{x}_i)$. Now the expectation value is being taken w.r.t all possible fields $\xi_i$ (and not the position $\vec{x}_i$) and hence the derivatives w.r.t  $i$ can be pulled out of the expectation value to get, $\langle \prod _{i=1}^n \delta _i \rangle \sim  \prod_{i=1}^n \nabla _i  \langle \prod_{i=1}^n \varphi_i \rangle $.\\
~\\
Now by the theorem about cumulants proven in appendix (\ref{fnl}) we have that only for $n=3$, does $\langle \varphi^n \rangle_c$ have a term which is linear in $f_{NL}$. This is the crucial observation which immediately cuts down our search to looking for only a finite number of shapes of product of cumulants if we want to keep track of terms linear order in $f_{NL}$.\\
~\\
In the cumulant expansion of any correlation function what occurs are product of all possible ways of partitioning the set of fields in the correlations. From the result above it follows that only $3$-point cumulants of $\delta$ $\langle \delta_i \delta _j \delta_k \rangle$ or a $2-point$ cumulant of a $\delta$ and a $y$ $\langle \delta_i y_j\rangle$ can give a $O(f_{NL})$ contribution. Further at $0^{th}$ order in $f_{NL}$ one can only have terms which are products of $2-$point functions. Also any of these $3-$point cumulants can come multiplied with an arbitrary number of $2-$point functions of the $\delta$ without altering the count of number of powers of $f_{NL}$ that they contribute. So in any of the correlation functions occurring in the expression (\ref{bin}) for $\langle \xi (\vert  \vec{x}_i - \vec{x}_j \vert_{(0<i<j=2,..,n)}) \rangle$ , to account for all terms of $O(f_{NL})$ we need to keep track of only the following kinds of product of cumulants, 

\begin{align} \label{shape}
\notag \prod _{(i,j)} \langle \delta_i \delta_j  \rangle \\
\notag \langle \delta_i \delta_j \delta _k \rangle \prod_{(p,q)} \langle \delta_p \delta_q \rangle \\
\langle \delta_i y_j \rangle \prod_{(p,q)} \langle \delta_p \delta_q \rangle
\end{align}

The first of the above three gives a $0^{th}$ order in $f_{NL}$ contribution and the later two give a linear order in $f_{NL}$ contribution. It is obvious that among the two kinds of correlations to be tracked in the R.H.S of (\ref{bin}) the first of the above kind of terms can occur in the cumulant expansion of a correlation with an even (say $2N$ for all $N \in \mathbb{W}$) number of $\delta$s, the second type can occur in correlations with $2M+3$ number of $\delta$s for all $M \in \mathbb{W}$ and the third type can occur in correlations with $2A+1$ number of $\delta$s and $1$ $y$ for all $A \in \mathbb{W}$. (..here I use $\mathbb{W}$ to denote the set of whole numbers, $\{0,1,2,3...\}$..) Hence these are the only three kinds of correlations among all the $\langle  \prod _{i=1}^n \delta _i ^{n_i} y_i^{m_i} \rangle$ and  $\langle \delta ^{n} y^{m} \rangle$ that need to be accounted for.

\subsection{The $O(f_{NL})$ terms in the denominator of  $1+\langle \xi (\vert  \vec{x}_i - \vec{x}_j \vert_{(0<i<j=2,..,n)}) \rangle$} 

By the argument above the terms contributing at linear or lower order in $f_{NL}$ come from the correlations in the denominator of the form $\langle \delta ^{2N} \rangle$, $\langle \delta ^{2N+3} \rangle$ and $\langle y\delta ^{2N+1} \rangle$ for every value of $N \in \{0,1,2,3..\}$. One sees that the the later of the two forms can only contribute terms which are linear order in $f_{NL}$ whereas the first type will produce terms at $0^{th}$order in $f_{NL}$. Doing the combinatorics of the coefficients in the cumulant expansion one sees that the relevant  terms in each type are (indicated with a right arrow),

\begin{align} \label{denop}
\notag \langle \delta ^{2N} \rangle  &\longrightarrow (2N)!\langle \delta ^2\rangle ^N\\
\notag \langle \delta ^{2N+3}\rangle &\longrightarrow ^{(2N+3)}C_3 (2N)! \langle \delta^3 \rangle \langle \delta^2 \rangle^N\\
\langle \delta ^{2N+1} y \rangle &\longrightarrow (2N+1)(2N)!\langle \delta y \rangle \langle \delta^2 \rangle ^N
\end{align}

So putting back the relevant factors of $c_{nm}$, the denominator of  $1+\langle \xi (\vert  \vec{x}_i - \vec{x}_j \vert_{(0<i<j=2,...,n)}) \rangle$ to linear order in $f_{NL}$ is given by,

\begin{align}\label{deno}
 \left [ \sum_{N=0}^{\infty} \left \{ \frac {c_{2N,0}}{(2N)!} (2N)! \langle \delta ^2\rangle ^N + \frac{c_{2N+3,0}}{(2N+3)!}^{(2N+3)}C_3 (2N)! \langle \delta^3 \rangle \langle \delta^2 \rangle^N + \frac{c_{2N+1,1}}{(2N+1)!}(2N+1)(2N)!\langle \delta y \rangle \langle \delta^2 \rangle ^N   \right \}  \right ]^n 
\end{align}

\subsection{The $O(f_{NL})$ terms in the numerator of  $1+\langle \xi (\vert  \vec{x}_i - \vec{x}_j \vert_{(0<i<j=2,...,n)}) \rangle$}

Among the correlations occurring in the numerator (which are of the form, $\langle \prod_{i=1}^n \delta_i ^{n_i} y_i ^{m_i} \rangle$) we look for product of cumulants of the shape (\ref{shape}) and we note that the following three ways are possible for all $N \in \mathbb{W}$,

\begin{align}\label{nump}
\notag \langle \prod_{i=1}^n \delta_i ^{N_i} \rangle \text{ with } \sum_{i=1}^n N_i = 2N &\longrightarrow \prod_{(p,q)} \langle \delta_p \delta_q \rangle \\
\notag \langle \prod_{i=1}^n \delta_i ^{N_i} \rangle \text{ with } \sum_{i=1}^n N_i = 2N+3 &\longrightarrow \langle \delta_i \delta_j \delta _k \rangle \prod_{(p,q)} \langle \delta_p \delta_q \rangle \\
\langle y_j  \prod_{i=1}^n \delta_i ^{N_i} \rangle \text{ with } \sum_{i=1}^n N_i = 2N+1 &\longrightarrow \langle \delta_i y_j \rangle \prod_{(p,q)} \langle \delta_p \delta_q \rangle
\end{align}

Now one needs to recognize that for any of the forms on the L.H.S of the arrow there are multiple ways in which its cumulant expansion can produce terms of the form as on the right of the arrow. These possible ways of splitting are kept track of by the same tableau imagination as in appendix (\ref{nzl}). Since in any of the three possible cases listed above the partitioning is among the occurrences of $\delta$ we need to imagine now only a $\delta$ tableau.\\
~\\
So define a tableau $T_{N_{1\leq i \leq n}}$ such that it has $n$ rows (corresponding to the  $n-$points of the tracer correlation) such that the $i^{th}$ row has $N_i$ boxes. Now one sees that the three possibilities of (\ref{nump}) are accounted for in the following way,

\begin{enumerate}
\item The first possibility of (\ref{nump}) occurs in those terms of the cumulant expansion of $\langle \prod_{i=1}^n \delta_i ^{N_i} \rangle$ which come from those partitions of the tableau $T_{N_{1\leq i \leq n}}$ (with $\sum_{i=1}^{n} N_i = 2N$)  which have only parts of size $2$. Let such partitions be denoted by $\rho_{(2)N}$. For any part $B \in \rho_{(2)N}$ let $B_1$ and $B_2$ be the two elements in it. Restoring the bare bias coefficients of (\ref{bin}) these contributions come as,

\begin{align}\label{num1}
\sum_{N=0}^{\infty} \sum_{ \substack { n_{1\leq i \leq n =0} \\
                                         \sum_i n_i = 2N} }^{\infty} \prod_{i=1}^n \left [ \frac{c_{n_i,0}} {n_i!} \right] \sum _{ \rho_{(2)N} \text{ of } T_{N_{1\leq i \leq n}}} \prod _{B \in \rho_{(2)N}} \langle \delta_{B_1} \delta_{B_2} \rangle 
\end{align}

\item The second possibility of (\ref{nump}) occurs in those terms of the cumulant expansion of  $\langle \prod_{i=1}^n \delta_i ^{N_i} \rangle$ which come from those partitions of the tableau $T_{N_{1\leq i \leq n}}$ (with $\sum_{i=1}^{n} N_i = 2N+3$) which have one part of size $3$ and all other parts of size $2$. Let such partitions be denoted by $\rho_{3+(2)N}$. For any such partition let the subscript $1,2,3$ denote the elements from the size $3$ part and the $B$ run over all other size $2$ parts with $B_1$ and $B_2$ denoting the two elements in it. Restoring the bare bias coefficients of (\ref{bin}) these contributions come as,

\begin{align}\label{num2}
\sum_{N=0}^{\infty} \sum_{ \substack { n_{1\leq i \leq n =0} \\
                                         \sum_i n_i = 2N+3} }^{\infty} \prod_{i=1}^n \left [ \frac{c_{n_i,0}} {n_i!} \right] \sum _{ \rho_{3+(2)N}\text{ of }T_{N_{1\leq i \leq n}}}   \langle \delta_1 \delta_2 \delta_3  \rangle \prod _{B \in \rho_{3+(2)N}} \langle \delta_{B_1} \delta_{B_2} \rangle 
\end{align}

\item The third possibility of (\ref{nump}) occurs in those terms of the cumulant expansion of  $\langle y_j \prod_{i=1}^n \delta_i ^{N_i} \rangle$ which come from those partitions of the tableau $T_{N_{1\leq i \leq n}}$ (with $\sum_{i=1}^{n} N_i = 2N+3$) which have one part of size $1$ and all other parts of size $2$. Let such partitions be denoted by $\rho_{1+(2)N}$. That singleton part combines with the $y_j$ to give the $\langle \delta_i y_j\rangle$ term. For any such partition let the subscript $1$ denote that element from the size $1$ part and the $B$ run over all other size $2$ parts with $B_1$ and $B_2$ denoting the two elements in it. Restoring the bare bias coefficients of (\ref{bin}) these contributions come as,

\begin{align}\label{num3}
\sum_{N=0}^{\infty} \sum_{ \substack { n_{1\leq i \leq n =0} \\
                                         \sum_i n_i = 2N+1} }^{\infty} \sum_{j=1}^n c_{0,1} \prod_{i=1}^n \left [ \frac{c_{n_i,0}} {n_i!} \right] \sum _{ \rho_{1+(2)N} \text{ of }  T_{N_{1\leq i \leq n}}}   \langle \delta_1 y_j  \rangle \prod _{B \in \rho_{1+(2)N}} \langle \delta_{B_1} \delta_{B_2} \rangle 
\end{align}

Here there is an extra sum over $j$ to account for the fact that the $y$ contribution can come from any of the $n$ tracer points among which the correlation is being calculated.
\end{enumerate}  

\newpage 

\subsection {The final answer} 

Let us call the expression in (\ref{num1}) as $A$ and the sum of  (\ref{num2}) and (\ref{num3}) as $B$. Let the first term in (\ref{deno}) be $C$ and the sum of its last two terms be $D$. Then $B,D$ are $O(f_{NL})$ and the rest are $0^{th}$ order. Then schematically we have, 

\begin{align}\label{schemf}
1+\langle \xi (\vert  \vec{x}_i - \vec{x}_j \vert_{(0<i<j=2,...,n)}) \rangle = \frac{ A + B }{(C + D)^n}
\end{align}

here $B$ and $D$ being $O(f_{NL})$ one can binomally expand the denominator to linear order and multiply that to the numerator and ignoring the $O(f_{NL}^2)$ terms of $BD$ one has,

\begin{align}\label{final}
1+\langle \xi (\vert  \vec{x}_i - \vec{x}_j \vert_{(0<i<j=2,...,n)}) \rangle =  \frac{1}{C^n} [A + B - \frac{nAD}{C}] + O(f_{NL}^2)
\end{align}

where to repeat,

\begin{align}
\notag A &= \sum_{N=0}^{\infty} \sum_{ \substack { n_{1\leq i \leq n =0} \\ \sum_i n_i = 2N} }^{\infty} \prod_{i=1}^n \left [ \frac{c_{n_i,0}} {n_i!} \right] \sum _{ \rho_{(2)N} \text{ of } T_{N_{1\leq i \leq n}}} \prod _{B \in \rho_{(2)N}} \langle \delta_{B_1} \delta_{B_2} \rangle  \\
\notag B &= \sum_{N=0}^{\infty} \sum_{ \substack { n_{1\leq i \leq n =0} \\
                                         \sum_i n_i = 2N+3} }^{\infty} \prod_{i=1}^n \left [ \frac{c_{n_i,0}} {n_i!} \right] \sum _{ \rho_{3+(2)N}\text{ of }T_{N_{1\leq i \leq n}}}   \langle \delta_1 \delta_2 \delta_3  \rangle \prod _{B \in \rho_{3+(2)N}} \langle \delta_{B_1} \delta_{B_2} \rangle \\
\notag            &+  \sum_{N=0}^{\infty} \sum_{ \substack { n_{1\leq i \leq n =0} \\
                                         \sum_i n_i = 2N+1} }^{\infty} \sum_{j=1}^n c_{0,1} \prod_{i=1}^n \left [ \frac{c_{n_i,0}} {n_i!} \right] \sum _{ \rho_{1+(2)N} \text{ of }  T_{N_{1\leq i \leq n}}}   \langle \delta_1 y_j  \rangle \prod _{B \in \rho_{1+(2)N}} \langle \delta_{B_1} \delta_{B_2} \rangle\\ 
\notag C &=  \left [ \sum_{N=0}^{\infty} \left \{c_{2N,0} \langle \delta ^2\rangle ^N \right \} \right ] \\
\notag D &=  \left [ \sum_{N=0}^{\infty} \left \{ \frac{c_{2N+3,0}}{6} \langle \delta^3 \rangle \langle \delta^2 \rangle^N + c_{2N+1,1}\langle \delta y \rangle \langle \delta^2 \rangle ^N   \right \}  \right ] \\
\end{align}

\newpage

\section {Conclusion}

We hope that the combinatorial technologies introduced in this work shall find further uses in the theory of understanding astrophysical correlations. An important question that immediately suggests itself for future work is to try to extend the renormalization result of \ref{renbiasuni} to the case of multivariate biasing. Towards that goal or even independently it is likely to be interesting to understand if the results of \ref{fnl} can be extended to higher powers of $f_{NL}$. That would possibly help in being able to develop a perturbative expansion of the L.H.S of \ref{final} in powers of $f_{NL}$ beyond the linear order result obtained here. 

\section {Acknowledgments} 
I would like to thank Neal Dalal for introducing me to this subject and for the many helpful discussions during which this project was evolved. I would like to thank the Department of Astronomy at UIUC for letting me use their kitchen during the course of these discussions. Thanks are due to Fabian Schmidt for kindly agreeing to read the manuscript and for giving his many helpful suggestions. I would like to thank the Materials Research Laboratory (MRL) of UIUC in whose reading room and lounges most of this work was done.

\appendix

\section {Proof of decomposition of arbitrary correlations into ``nzl" and ``zl" correlations} \label{nzl} 

Let us do the proof for a slightly general case of having two fields say $\delta$ and $y$. We use the notation of $A_i$ to mean $A(\vec{x}_i)$ for any field $A = \delta$ or $A =y$.  So we want the required decomposition on the correlation, $\langle \prod_{i=1}^{n} \delta_i ^{n_i} y^{m_i}_i  \rangle$ for any set of positive integers $n_i$ and $m_i$.\\
Here we define the required tableau to have $2n$ rows, where the first $n$ of them are labeled by $\delta_i$ for $i=1,...,n$ and the next $n$ are labeled by $y_i$ for  $i=1,...,n$. Each of the $\delta_i$ row has $n_i$ boxes and each of the $y_i$ row has $m_i$ boxes. Then as earlier let $\rho$ run over all partitions of the tableau and $B$ label the parts of the partition. One imagines a $\rho$ as a set of sets whereby each of the element sets is a $B$. Further define, $n_{\delta_i} (B)$ and $n_{y_i}(B)$ as the number of elements in $B$ which come from the $\delta_i$ and $y_i$ row respectively. Then one has the obvious rewriting, 

\begin{align}
\langle \prod_{i=1}^{n} \delta_i ^{n_i} y^{m_i}_i  \rangle = \sum _ {\rho} \prod _{B \in \rho} \langle \prod_{i=1}^{n} \delta_i ^{n_{\delta_i}(B)} y_i ^{n_{y_i}(B)} \rangle_c 
\end{align} 

Now define $N_i(\rho)$ as the number of elements from the $\delta_i$ row which occur in some/any $B \in \rho$ s.t $B$ has elements only from the $\delta_i$ row. So one can write as an equation, 

\begin{align}
N_i (\rho) = \sum _{\substack {B \in \rho \text{ s.t.} \\ 
      n_{\delta_{\!j}}(B) = 0 \\
      \forall j \neq i}} 
n_{\delta_i} (B) 
\end{align}

Similarly define $M_i (\rho)$ for the corresponding $y_i$ whereby one replaces the $\delta_i$ by $y_i$ in the above definition.  Now one notes that $ 0 \leq N_i(\rho) \leq n_i$ and $0 \leq M_i(\rho) \leq m_i$ and that valid tuples of values for $\{ N_i(\rho),M_i(\rho), i=1,...,n\}$ defines a partition of the set of partitions of the tableau. So any $\rho$ falls into precisely one such equivalence class defined by its tuple, $\{ N_i(\rho),M_i(\rho), i=1,...,n\}$. So one can rearrange the above expansion of the correlation function as a sum over the equivalence classes to give, 

\begin{align}
\langle \prod_{i=1}^{n} \delta_i ^{n_i} y^{m_i}_i  \rangle = \sum _ {a_{i=1...n} = 0}^{n_{i=1..n}} \sum_{b_{i=1...n} = 0} ^{m_{i=1...n}} \sum _{\substack{ \rho \text{ s.t.}\\ 
      N_i(\rho) = a_i \\
      \text{and }M_i(\rho) = b_i} }
\prod _{B \in \rho} 
\biggl\langle 
   \prod_{i=1}^{n} \delta_i ^{n_{\delta_i}(B)} 
   y_i ^{n_{y_i}(B)} 
\biggr\rangle_{\!c}
\end{align} 
  
Now for any term in the sum there are $^{n_i}C_{a_i}$ ways in which these $a_i$ elements of the $\delta_i$ row can be chosen and similarly there are $^{m_i}C_{b_i}$ ways these $b_i$ elements can be chosen from the $y_i$ row. \\
~\\
So the remaining  $n_i - a_i$ elements from the row $\delta_i$ and the $m_i - a_i$ elements from the row $y_i$ necessarily come in products of connected correlations such that each connected correlation has contribution from at least two rows and hence a contribution from at least two points. We naturally denote this part of the correlation as, $\langle \prod_{i=1}^{n} \delta_i ^{n_i - a_i} y_i ^{m_i - b_i} \rangle_{nzl}$.\\
~\\
Define $\rho_{\{p_i\}_{i=1,..,n}}$ to be the partitions of a tableau which has $n$ rows and there are $p_i$ boxes in each row for $i=1$ to $i=n$. Then we can rewrite the above as,  
 
\begin{align}
\biggl\langle \prod_{i=1}^{n} \delta_i ^{n_i} y^{m_i}_i \biggr\rangle 
 &= 
\sum _ {a_{i=1,\dots, n} = 0}^{n_{i=1,\dots, n}} 
\sum_{b_{i=1,\dots, n} = 0} ^{m_{i=1,\dots, n}} 
\prod_{i=1}^{n} \Bigg[
  {}^{n_i}C_{a_i} \biggl(
    \sum _ {\rho_{\{a_{1\leq i \leq n}\}}} 
    \prod _{B \in \rho_{\{ a_{1 \leq i \leq n} \}}} 
    \bigl\langle \delta_i ^{\vert B \vert} \bigr\rangle _c 
  \biggr) \notag\\ 
 &\quad \times {}^{m_i}C_{b_i}
  \biggr( \sum _ {\rho_{\{b_{1\leq i\leq n}\}}} 
    \prod _{B \in \rho_{\{b_{1\leq i \leq n} \}}} 
    \langle y_i ^{\lvert B \rvert} \rangle_c 
  \biggr) \Bigg] 
  \biggl\langle 
    \prod_{i=1}^{n} \delta_i ^{n_i - a_i} y_i ^{m_i - b_i} 
  \biggr\rangle _{\text{nzl}}
\end{align}

One can accumulate the single point connected correlations into full correlations and we get the final desired decomposition as, 

\begin{align}\label{nzlfinal}
\langle \prod_{i=1}^{n} \delta_i ^{n_i} y^{m_i}_i  \rangle = \sum _ {a_{i=1...n} = 0}^{n_{i=1..n}} \sum_{b_{i=1...n} = 0} ^{m_{i=1...n}} \prod_{i=1}^{n} \left [ ^{n_i}C_{a_i} \langle \delta_i ^{a_i} \rangle  ^{m_i}C_{b_i} \langle y_i ^{b_i}\rangle ^ { }    \right ] \langle \prod_{i=1}^{n} \delta_i ^{n_i - a_i} y_i ^{m_i - b_i} \rangle _{nzl} 
\end{align} 

The above argument obviously naturally generalizes to arbitrary number of fields.

\newpage

\section{ Vanishing of $O(f_{NL})$ terms for $4$ and higher point cumulants} \label{fnl} 

Here we want to show that for $n\geq 4$ there are no linear or lower order terms in $f_{NL}$ when $\langle (\phi + f_{NL}(\phi^2 - \langle \phi^2 \rangle)^n \rangle_c$ is written as a polynomial in $f_{NL}$. Here $\phi$ is a Gaussian/free field. (..one can heuristically think of $\phi$ as the inflaton field in the interacting picture..)\\
Since cumulants are not additive in general there is no direct way of calculating a cumulant. One almost direct way is to write it as a sum of products of correlations using the expression, 
\begin{align}
\langle Y^n \rangle_c = \sum _{\rho} (\vert \rho \vert - 1)!(-1)^{\vert \rho\vert -1}\prod_{B \in \rho} \langle Y^{\vert B\vert}\rangle
\end{align}

where $\rho$ runs over all possible partitions of the set $\{1,2,...,n \}$ and $B$ are the parts in any partition, $\vert \rho \vert$ is the number of parts in the partition $\rho$ and $\vert B\vert$ is the number of elements in the subset $B$. \\
~\\
Substituting $Y =  \phi + f_{NL}(\phi^2 - \langle \phi^2 \rangle)$ into the above one can check that for $n\geq 4$ there are no linear or lower order terms in $f_{NL}$. We note the following special cases that till quadratic order in $f_{NL}$ starts one has, $\langle Y \rangle = 0$ (one has shifted the ``vacuum" such that $\langle \phi \rangle =0$), $\langle Y^2 \rangle_c \sim \langle \phi^2 \rangle$, $\langle Y^3 \rangle_c \sim 6f_{NL}\langle \phi^2 \rangle^2$\\
~\\
Having checked the statement to be true for $n=4$ we can now start an induction using the recursion relation,

\begin{align}
\langle Y^n \rangle _c = \langle Y^n \rangle - \sum _{p=1}^{n-1}\text{ }^{n-1}C_{p-1} \langle Y^p\rangle_c \langle Y^{n-p}\rangle 
\end{align}

Let us assume that till some $m\geq 4$ one has proven that there are no linear or lower order terms in $\langle Y^n \rangle_c$. Then we want to show that the same holds for $m+1$. \\
~\\
For $\langle Y^{m+1}\rangle_c$ the sum on the RHS goes till $m$ and by the induction hypothesis any $\langle Y^p \rangle_c$ with $m>p>3$ will surely give at least a $f_{NL}^2$ contribution and hence is not relevant to our checking. Since we want to check for the vanishing of only linear or lower order terms in $f_{NL}$, on the RHS we need to keep only those $\langle Y^p \rangle_c$ where $p=2$ and $3$.\\
~\\
So we want to evaluate the linear or lower order in $f_{NL}$ terms in the RHS of the expression,  

\begin{align}
\langle Y^{m+1}\rangle_c = \langle Y^{m+1}\rangle - \sum _{p=1} ^m\text{ }^{m}C_{p-1} \langle Y^p\rangle_c \langle Y^{m+1-p}\rangle 
\end{align}

Now truncating the RHS of the above to whatever is potentially linear or lower order in $f_{NL}$ we get,

\begin{align}
\notag \langle (\phi + f_{NL}(\phi^2 - \langle \phi^2 \rangle)^{m+1} \rangle_c  &\sim  \langle \phi^{m+1} \rangle + (m+1)f_{NL}\{\langle \phi^{m+2}\rangle - \langle \phi^m\rangle \langle \phi^2 \rangle  \} \\
\notag &- m \langle \phi^2 \rangle \{ \langle \phi^{m-1} \rangle + f_{NL}(m-1) ( \langle \phi^m \rangle - \langle \phi^{m-2} \rangle \langle \phi^2 \rangle  )  \}\\
&- \frac{m(m-1)}{2} 6f_{NL} \langle \phi^2 \rangle ^2 \{ \langle \phi^{m-2}\rangle \}
\end{align}

Now one evaluates the above expression for two different cases once when $m$ is even and once when its odd and one finds that this vanishes. (\dots in each case one uses the identities for the Gaussian field that $\langle \phi ^{odd} \rangle = 0$ and $\langle \phi^{2q} \rangle = \frac{(2q)!}{q!2^q} \langle \phi^2 \rangle ^q$\dots).\\
~\\
Hence one has shown that  $\langle (\phi + f_{NL}(\phi^2 - \langle \phi^2 \rangle)^n \rangle_c$ has no linear order term in $f_{NL}$ for $n \geq 4$. 

\newpage

\section {Is there a notion of ``universality" in doing the ``nzl" decomposition for multivariate biasing?  }\label{universal} 

We try to push through the above technology for the case of having a two variable biasing and as of now we find that it does not go through as cleanly. But we still do get a splitting which can be seen as giving us a series expansion of the tracer $n-$point functions in terms of ``nzl" correlations with coefficients in terms of an ``universal" function.\\
~\\
Let the function $F_h$ be now allowed to depend on another variable say ``y"  as $F_h(\delta,y;\vec{x})$ and we use the notation of $F^{(n,m)}_h(0,0;\vec{x})$ to mean the evaluation at $\delta = y = 0$ the $n-$fold and $m-$fold derivative of $F_h$ w.r.t $\delta$ and $y$ respectively.  Then in terms of the ``nzl" correlations we write the expression for the $n-$point function in the case of bivariate biasing. 

(..we define $N_i = n_i - a_i$ and $M_i = m_i - b_i$  and whenever not mentioned the ``i" index on the RHS goes from $i=1$ to $i=n$..) 

\begin{align}
\langle \xi_h ( \vert \vec{x}_i - \vec{x}_j \vert ) \rangle  = -1 + \frac{A}{B} \\
\end{align}

where, 

\begin{align}
\notag A = \sum _ {\{ N_i = 0, M_i = 0\} }^{\infty}  &\left [  \sum_{\{n_i = N_i, m_i = M_i \}}^{\infty}   \langle \prod_{i=1}^n \frac{ F_h ^{(n_i,m_i)} (0,0;\vec{x}_i) }{n_i ! m_i !}    \rangle \prod _{i=1}^n \left [  ^{n_i}C_{n_i - N_i} \text{ }^{m_i}C_{m_i-M_i} \langle \delta_i ^{n_i - N_i}\rangle \langle y_i ^{m_i - M_i} \rangle \right ]   \right]\\
 &\langle \prod_{i=1}^n \delta_i ^{N_i} y_i ^{M_i} \rangle_{nzl}  
\end{align}

\begin{align}
B = { \prod_{i=1}^{n} \left [ \sum_{\{  n_i = 0, m_i = 0, i = 1,\dots,n \} }^{\infty} \frac{\langle F_h ^{(n_i, m_i)} ( 0,0;\vec{x}_i)  \rangle  } {n_i ! m_i ! } \langle \delta _i ^{n_i} y_i ^{m_i} \rangle  \right ]  } 
\end{align}

With the same two assumptions (\ref{Fas1} and \ref{Fas2}) on $F$ as earlier the above can be easily seen to rearrange to, 

\begin{align} 
\notag \langle \xi_h ( \vert \vec{x}_i - \vec{x}_j \vert ) \rangle = \sum _ {\{ N_i =1, M_i =1\}}^{\infty} \prod_{i=1}^{n} &\left [ \frac { \sum_{n_i = N_i, m_i = M_i}^{\infty}  \langle \frac{ F_h ^{(n_i,m_i)} (0,0;\vec{x}_i) }{n_i ! m_i !}    \rangle   ^{n_i}C_{n_i - N_i} \text{ } ^{m_i}C_{m_i-M_i} \langle \delta_i ^{n_i - N_i}\rangle \langle y_i ^{m_i - M_i} \rangle }  {\sum_{n_i =0, m_i =0 }^{\infty} \frac{\langle F_h ^{(n_i, m_i)} ( 0,0;\vec{x}_i)  \rangle  } {n_i ! m_i ! } \langle \delta _i ^{n_i} y_i ^{m_i} \rangle  }   \right ] \\
&\langle \prod_{i=1}^{n} \delta_i ^{N_i} y_i ^{M_i}  \rangle _{nzl} 
\end{align}

One can define the function $f_i$ as, (...the subscript of $i$ seems to denote that it is a different function at every spatial point of $\vec{x}_i$ but clearly it is not by the assumptions of homogeneity...)  

\begin{align}
f_i (N,M) = \sum _ {n_i = N_i, m_i = M_i} ^{\infty} \langle F^{(n_i, m_i)}(0,0;\vec{x}_i)\rangle \frac{\langle \delta_i ^{n_i - N}\rangle} {(n_i - N)!} \frac {\langle y_i ^{m_i - M} \rangle  } {(m_i - M)!}  
\end{align} 

Then in terms of the above function one can write the $n-$point function as, 

\begin{align}
\langle \xi_h ( \vert \vec{x}_i - \vec{x}_j \vert ) \rangle = \frac{ \sum _ {N_i =1, M_i =1} ^{\infty} \langle \prod_{i=1}^{n} \delta_i ^{N_i} y_i ^{M_i} \rangle_{nzl} \left [  \prod_{i=1}^{n} f_i (N_i, M_i) \right ]   } { \prod_{i=1} ^{n} \left [ \sum_{n_i =0, m_i =0 }^{\infty} \frac{\langle F_h ^{(n_i, m_i)} ( 0,0;\vec{x}_i)  \rangle  } {n_i ! m_i ! } \langle \delta _i ^{n_i} y_i ^{m_i} \rangle \right ]  }    
\end{align}

In the denominator one has for each point $\vec{x}_i$ the sum over terms of the form, $ \langle \delta _i ^{n_i} y_i ^{m_i} \rangle$. On these terms too one can do a nzl kind of decomposition to separate out the potentially divergent self-correlations of the fields at any one point. Doing that here one doesn't get exactly the same thing as one called as ``nzl" correlations but something more restrictive than that which can be denoted as, $\langle \delta_i ^{N_i} y_i ^{M_i}\rangle_{*}$. These $*$ correlations when expanded as sum of products of connected correlations gets only such connected correlations to contribute where there is at least one $\delta_i$  and one $y_i$.  This is more restrictive than the ``nzl" correlations because in an ``nzl" correlation it is possible for such a connected correlation to appear which has only $\delta$s or only $y$s (though all the $\delta$s and the $y$s can't be from the same point).\\
~\\
One curiosity for doing the above $*$ decomposition is that it reproduces the $f$ function in the denominator since, 

\begin{align}
\langle \xi_h ( \vert \vec{x}_i - \vec{x}_j \vert ) \rangle = \frac{ \sum _ {N_i =1, M_i =1} ^{\infty} \langle \prod_{i=1}^{n} \delta_i ^{N_i} y_i ^{M_i} \rangle_{nzl} \left [  \prod_{i=1}^{n} f_i (N_i, M_i) \right ]   } {\prod_{i=1} ^{n} \left [ \sum_{n_i =0, m_i =0 }^{\infty} \langle \delta_i ^{n_i} y_i ^{m_i} \rangle _{*} f_i (n_i,m_i) \right ]  }   
\end{align}

Though the separation is not as strong as in the single variable case (\ref{renbiasuni}) but one can think of the above as a power series expansion in the asymptotically decaying contribution (the ``nzl"s) with the functionally separated coefficients which absorb the potentially divergent and coarse graining scale dependent contributions (into the $f_i$s).  By homogeneity it follows that the $f$ functions are not dependent on the position $\vec{x}_i$ and hence further justifying the tag ``universal".

\newpage

\section { Comments on PBS bias parameters in the case of primordial local non-Gaussianity} \label{y} 

Here we review and make some comments about the framework setup in \cite{FDV}. 
One way to model how non-Gaussianity in the primordial fluctuations affect biasing is to imagine that it makes small scale fluctuations also matter. We imagine that this is effected through fluctuations at some scale say $R*$ getting important where $R*$ is smaller than the scale at which the $\delta$ used till now was being smoothed. Then one defines a quantity $\delta_s$ which measures the difference in fluctuations in the two scales as $\delta_s = \delta* - \delta$, where $\delta*$ is matter fluctuations smoothed at a scale $R*$. Inspired by what combination seems to appear in discussions of universal mass functions one wants to parameterize the dependence of $n_h$ through the variable $y^{*}(\vec{x})  = \frac{1}{2} \left ( \frac {\delta_s ^2 (\vec{x}) } { \sigma_s^2 }  - 1 \right )$, where $\sigma_s ^2 = \langle \delta_s^2 (\vec{x})\rangle$.\\
~\\
One pauses to note that one hasn't defined the conventional $f_{NL}$ in the above formalism. In the usual way of encoding non-Gaussianity one says that the matter fluctuations in the presence of non-Gaussianity $\delta_{NG}$ and the Gaussian one $\delta$ and the primordial gravitational potential ($\phi_p$) are related as, $\delta_{NG}^2 = \delta^2 [ 1 + 4f_{NL} \phi_p] + O(f_{NL}^2)$. It turns out that the results in the two formalisms can be related (though not totally equivalent) by imagining a relationship like, $y = \frac{1}{2}(\frac{\delta_{NG}^2} {\delta^2 } -1) $. A key point of the current formalism is to to avoid putting in such an explicit equation of $\delta_{NG}$ in terms of $f_{NL}$ but to let the arbitrary statistic $\delta$ and $\delta_s$ be related to the primordial potential through whatever comes from the gravitation model. Like (using the conventions of Dodelson's book) one has in a general FRW universe that at linear order the $k^{th}$ momentum modes of these two things are related as, $ \frac{\phi_{kp}}{\delta_k} = \frac{5  (aH)^2\Omega} {3k^2 T(k) \frac{D(a)}{a}}$. \\
~\\
Coming back to the current formalism one wants to ask here as to how does $n_h$ change depending on a scaling of matter fluctuations as $\delta \rightarrow (1+\epsilon) \delta$ which also changes $y^{*}$ as $y^{*} \rightarrow (1+\epsilon)^2y^{*} + \left ( \epsilon + \frac{\epsilon^2}{2}  \right)$. Then one as before assumes the existence of a function $F_h$ such that $n_h (\vec{x}) = F_{h}(\delta, y^{*})$  and power series expands this function in two variables. One defines $c_{nm} = \frac{1}{\langle F_h(0,0) \rangle} F^{(n,m)}_h(\delta,y^{*})\vert _ {\delta = y^{*} = 0}$. It also assumes a certain lack of correlation as in,

\begin{align}
\langle  F^{(n,m)}_h(\delta,y^{*})\vert _ {\delta = y^{*} = 0}  \delta ^n y^{*m} \rangle =  \langle  F^{(n,m)}_h(\delta,y^{*})\vert _ {\delta = y^{*} = 0} \rangle \langle  \delta ^n y^{*m} \rangle
\end{align} 

Then including the previous dependency on $D$ too one has in this case, 

\begin{align}
\langle n_h (\vec{x},D,\epsilon) \rangle = \langle F_h (\vec{x},0,0) \rangle \sum _{n,m =0} ^{\infty} \frac{c_{nm}} {n! m!} \langle [ (1+\epsilon)\delta(\vec{x}) + D]^n \left [(1+\epsilon)^2y^{*}(\vec{x}) + \left (  \epsilon + \frac{\epsilon^2}{2} \right ) \right ]^m \rangle 
\end{align} 

It is understood in the above writing that the $\vec{x}$ dependencies don't remain after taking the expectation values and often they will be dropped within expectation values. But it is important to remember this dependence for $n_h$ since eventually one would want to calculate the $n-point$ function of the tracers where the expectation will be taken of the product of $n_h$ at $n$ different points. \\
~\\
One wants to define the bi-variate PBS bias parameters ($b_{NM}$) in this formalism as,  

\begin{align}
b_{NM} = \frac{1}{\langle n_h (\vec{x},D=0,\epsilon=0) \rangle} \frac { \partial ^{N+M} \langle n_h (D,\epsilon) \rangle  } {\partial D^N \partial \epsilon ^M } \vert _ {D = \epsilon = 0}
\end{align}

Before taking the derivatives in the above equation it is helpful to binomially expand the definition of $\langle n_h (D,\epsilon) \rangle$ to get,  

\begin{align}
\notag &\langle n_h (\vec{x},D,\epsilon) \rangle =\\
&\tilde{\sum}_{n,m=0}^{\infty} \frac{\langle F^{(n,m)}(\vec{x},0,0) \rangle} {n! m!} \tilde{\sum}_{p=0}^{n} \tilde{\sum} _{q=0}^{m}\text{ }^{n}C_p\text{ }^{m}C_q (1+\epsilon)^{p+2(m-q)} \left (\epsilon + \frac{\epsilon^2}{2} \right )^q D^{n-p} \langle \delta^p y^{*(m-q)} \rangle 
\end{align} 

A $\tilde{}$ has been put on all the $\sum$ to indicate that the $4$ sums are not free sums but can go over only those values of $n,m,p,q$ such that the summand makes sense i.e it satisfies the following validity inequalities, $n\geq p, m\geq q, p+2(m-q), q, n-p, p, m-q \geq 0$. Its not worth trying to restrict the sum in some complicated way to ensure these inequalities are automatically satisfied but in any particular case if necessary one would check or explicitly implement them.\\
~\\
Before taking the derivatives one would further expand the $\epsilon$ factors to get, 

\begin{align}
(1+\epsilon)^{p+2(m-q)} (\epsilon + \frac{\epsilon^2}{2})^q = \sum _ {a =0} ^{p+2(m-q)} \sum _{b=0}^{q} \frac{1}{2^b} {}^{p+2(m-q)}C_a {}^{q}C_b \epsilon ^{a+b+q}
\end{align}

After doing the requisite differentiation one has the following expression for the bivariate bias-parameters, 

\begin{align} 
b_{NM} = \frac{N! M!}{\langle n_h (\vec{x},D=0,\epsilon=0) \rangle 2^M} \sum _ {n = N,m=0}^{\infty}   \frac{\langle F^{(n,m)}(\vec{x},0,0) \rangle} {n! m!} ^{n}C_N \sum _{q=0} ^{m} \sum_{a=0}^{(n-N) + 2(m-q)} \\
 2^{a+q}\text{ }^{m}C_q  ^{(n-N)+2(m-q)}C_a\text{ }^{q}C_{M-(a+q)} \langle  \delta ^{n-N} y^{m-q} \rangle 
\end{align}

One notes that in the above expression there are no $\tilde{}$ on the $\sum$. Here everything is a free sum and the validity inequalities have all been accounted for.\\
~\\

As a special case of the above one notes the following,

\begin{align}
b_{N0} = \frac{1}{\langle n_h (\vec{x},D=0,\epsilon=0) \rangle } \sum_{n=N, m=0} ^{\infty}  \frac{\langle F^{(n,m)}(\vec{x},0,0) \rangle} {(n-N)! m!} \langle \delta (\vec{x})^{n-N} y^{*m} \rangle 
\end{align}

One notes that this $b_{N0} \neq b_N$ (\ref{bn}). One notes in the same vein of comparing between the univariate and the bivariate case that, $n_h (D=0,\epsilon =0) \neq n_h (D=0)$.

%

\bibliographystyle{unsrt}

\begin{thebibliography}{99}

\bibitem{FDV} Fabian Schmidt, Donghui Jeong, Vincent Desjacques, ``Peak-Background Split, Renormalization and Galaxy Clustering", \url{http://arxiv.org/abs/1212.0868v2}

\bibitem{NGLS} Vincent Desjacques, Uros Seljak, ``Primordial non-Gaussianity in the Large Scale Structure of the Universe", \url{http://arxiv.org/abs/1006.4763v1}, Anze Slosar, Christopher Hirata, Uros Seljak, Shirley Ho, Nikhil Padmanabhan, ``Constraints on local primordial non-Gaussianity from large scale structure", \url{http://arxiv.org/abs/0805.3580v2}, Uros Seljak, ``Extracting primordial non-gaussianity without cosmic variance", \url{http://arxiv.org/abs/0807.1770v1}, Roman Scoccimarro, Emiliano Sefusatti, Matias Zaldarriaga, ``Probing Primordial Non-Gaussianity with Large-Scale Structure", \url{http://arxiv.org/abs/astro-ph/0312286}, N. Bartolo, S.Matarrese, A.Riotto, ``Signatures of Primordial Non-Gaussianity in the Large-Scale Structure of the Universe", \url{http://arxiv.org/abs/astro-ph/0501614.pdf}

\bibitem{WMAP9}G. Hinshaw, D. Larson, E. Komatsu, D. N. Spergel, C. L. Bennett, J. Dunkley, M. R. Nolta, M. Halpern, R. S. Hill, N. Odegard, L. Page, K. M. Smith, J. L. Weiland, B. Gold, N. Jarosik, A. Kogut, M. Limon, S. S. Meyer, G. S. Tucker, E. Wollack, E. L. Wright, ``Nine-Year Wilkinson Microwave Anisotropy Probe (WMAP) Observations: Cosmological Parameter Results",\url{http://arxiv.org/abs/1212.5226}

\bibitem{Dodelson} Scott Dodelson, ``Modern Cosmology" 

\bibitem{BF}R. H. Brandenberger and C. Vafa, Superstrings in the Early Universe," Nucl. Phys. B 316, 391 (1989);
A. Nayeri, R. H. Brandenberger and C. Vafa, Producing a scale-invariant spectrum of perturbations in a Hagedorn phase of string cosmology," Phys. Rev. Lett.97, 021302 (2006) \url{http://arxiv.org/abs/hep-th/0511140}; R. H. Brandenberger, A. Nayeri, S. P. Patil and C. Vafa, String gas cosmology and structure formation," Int. J.Mod. Phys. A 22, 3621 (2007) \url{http://arxiv.org/abs/hep-th/0608121}; R. H. Brandenberger, String Gas Cosmology," \url{http://arxiv.org/abs/0808.0746}

\bibitem{Sandip} Shamit Kachru, Renata Kallosh, Andrei Linde, Sandip P. Trivedi, ``de Sitter Vacua in String Theory", \url{http://arxiv.org/abs/hep-th/0301240 }, Shamit Kachru, Renata Kallosh, Andrei Linde, Juan Maldacena, Liam McAllister, Sandip P. Trivedi, ``Towards Inflation in String Theory", \url{http://arxiv.org/abs/hep-th/0308055}, Ishan Mata, Suvrat Raju, Sandip Trivedi, ``CMB from CFT", \url{http://arxiv.org/abs/1211.5482}

\bibitem{Creminelli} Clifford Cheung, Paolo Creminelli, A. Liam Fitzpatrick, Jared Kaplan, Leonardo Senatore, ``The Effective Field Theory of Inflation", \url{http://arxiv.org/abs/arXiv:0709.0293}, Clifford Cheung, A. Liam Fitzpatrick, Jared Kaplan, Leonardo Senatore, ``On the consistency relation of the 3-point function in single field inflation", \url{http://arxiv.org/abs/0709.0295}, Leonardo Senatore, Matias Zaldarriaga, ``A Naturally Large Four-Point Function in Single Field Inflation", \url{http://arxiv.org/abs/1004.1201  }, Leonardo Senatore, Matias Zaldarriaga ``The Effective Field Theory of Multifield Inflation", \url{http://arxiv.org/abs/1009.2093}

\bibitem {Randall} Lisa Randall, Martin Soljacic, Alan Guth, ``Supernatural Inflation: Inflation from Supersymmetry with No (Very) Small Parameters", \url{http://arxiv.org/abs/hep-ph/9512439}, Lisa Randall, Raman Sundrum, ``A Large Mass Hierarachy from a Small Extra Dimension", \url{http://arxiv.org/abs/hep-ph/9905221}, Nima Arkani-Hamed, Hsin-Chia Cheng, Paolo Creminelli, Lisa Randall, ``Extranatural Inflation", \url{http://arxiv.org/abs/hep-th/0301218}, Nima Arkani-Hamed, Hsin-Chia Cheng, Paolo Creminelli, Lisa Randall, ``Pseudonatural Inflation", \url{http://arxiv.org/abs/hep-th/0302034}, Nima Arkani-Hamed, Paolo Creminelli, Shinji Mukhoyama, Matias Zaldarriaga, ``Ghost Inflation", \url{http://arxiv.org/abs/hep-th/0312100}   

\bibitem {Khoury} Kurt Hinterbichler, Austin Joyce, Justin Khoury, ``Non-linear realizations of Conformal Symmetry and Effective Field Theory for the Pseudo-Conformal Universe", \url{http://arxiv.org/pdf/1202.6056.pdf }, Paolo Creminelli, Austin Joyce, Justin Khoury and Marko Simonovic, ``Consistency Relation for the Conformal Mechanism", \url{http://arxiv.org/pdf/1212.3329.pdf}

\bibitem{B} Daniel Baumann, ``TASI lectures on Inflation", \url{http://arxiv.org/pdf/0907.5424.pdf}

\bibitem {Jessie} Mark P. Hertzberg, Max Tegmark, Shamit Kachru, Jessie Shelton, Onur Ozcan, ``Searching for Inflation in Simple String Theory Models: An Astrophysical Perspective", \url{http://arxiv.org/abs/arXiv:0709.0002}


\bibitem{Sarah} Marilena LoVerde, Amber Miller, Sarah Shandera, Licia Verde, ``Effects of Scale-Dependent Non-Gaussianity on Cosmological Structures", \url{http://arxiv.org/abs/0711.4126v3.pdf}, Sarah Shandera, Neal Dalal, Dragan Huterer, ``A generalized local ansatz and its effect on halo bias", \url{http://arxiv.org/abs/1010.3722v2.pdf}

\bibitem{D} Vincent Desjacques, Donghui Jeong, Fabian Schmidt, ``Non-Gaussian Halo Bias Re-examined: Mass-dependent Amplitude from the Peak-Background Split and Thresholding", \url{http://arxiv.org/abs/1105.3628} Emiliano Sefusatti, Martin Crocce, Vincent Desjacques, ``The Halo Bispectrum in N-body Simulations with non-Gaussian Initial Conditions", \url{http://arxiv.org/abs/1111.6966}, Donghui Jeong, Fabian Schmidt, Christopher M. Hirata, ``Large-scale clustering of galaxies in general relativity", \url{http://arxiv.org/abs/1107.5427}, Vincent Desjacques, Jinn-Ouk Gong, Antonio Riotto, ``Non-Gaussian bias: insights from discrete density peaks", \url{http://arxiv.org/abs/1301.7437v2.pdf}

\bibitem{Baldauf} Tobias Baldauf, Uros Seljak, Leonardo Senatore, ``Primordial non-Gaussianity in the Bispectrum of the Halo Density Field", \url{http://arxiv.org/pdf/1011.1513v3.pdf}, Tobias Baldauf, Uros Seljak, Leonardo Senatore, Matias Zaldarriaga, ``Galaxy Bias and non-Linear Structure Formation in General Relativity", \url{http://arxiv.org/abs/1106.5507}, Tobias Baldauf, Uros Seljak, Vincent Desjacques, Patrick McDonald, ``Evidence for Quadratic Tidal Tensor Bias from the Halo Bispectrum", \url{http://arxiv.org/abs/1201.4827},   


\bibitem{others} Emiliano Sefusatti, Martin Crocce, Vincent Desjacques, ``The Halo Bispectrum in N-body Simulations with non-Gaussian Initial Conditions", \url{http://arxiv.org/abs/1111.6966}, Daniel Baumann, Simone Ferraro, Daniel Green, Kendrick M.Smith, ``Stochastic Bias from Non-Gaussian Initial Conditions", \url{http://arxiv.org/abs/1209.2173.pdf}, Ravi K.Sheth, Kwan Chuen Chan, Roman Scoccimarro, ``Non-local Lagrangian Bias", \url{http://arxiv.org/abs/1207.7117.pdf}, Donghui Jeong, Eiichiro Komatsu, ``Perturbation Theory Reloaded II:Non-linear Bias, Baryon Acoustic Oscillations and Millenium Simulation in Real Space", \url{http://arxiv.org/abs/0805.2632.pdf},  Donghui Jeong, Eiichiro Komatsu, ``Primordial non-Gaussianity, scale-dependent bias, and the bispectrum of galaxies", \url{http://arxiv.org/abs/0904.0497.pdf}
\end{thebibliography}

\end{document}